\documentclass[aps,epsf,prl,epsfig,twocolumn]{revtex4}
\def\be{\begin{equation}}
\def\ee{\end{equation}}
\def\bea{\begin{eqnarray}}          
\def\eea{\end{eqnarray}}
\def\bi{\begin{itemize}}
\def\ei{\end{itemize}}

\begin{document}

\title{  Images of a Bose-Einstein condensate 
            in position and momentum space        
\footnote{ Talk given at the Laser Physics Workshop, 
                 July 2005, Kyoto, Japan}                       }

\author{ Jacek Dziarmaga and Krzysztof Sacha }

\address{ Institute of Physics and Centre for Complex Systems,
          Jagiellonian University,
          Reymonta 4, 30-059 Krak\'ow, Poland  }

\date{ December 08, 2005 }

\begin{abstract}
In the Bogoliubov theory a condensate initially prepared 
in its ground state described by stationary Bogoliubov
vacuum and later perturbed by a time-dependent potential
or interaction strength evolves into a time-dependent
excited state which is dynamical Bogoliubov vacuum.
The dynamical vacuum has a simple diagonal form in a
time-dependent orthonormal basis of single particle modes.
This diagonal representation leads to a gaussian probability
distribution for possible outcomes of density measurements 
in position and momentum space. In these notes we also discuss
relations with the $U(1)$ symmetry breaking version of
the Bogoliubov theory and give two equivalent gaussian
integral representations of the dynamical vacuum state.
\end{abstract}
\maketitle

\section{ Introduction }

It has been established in a number of papers \cite{Fock,phase} that
a density measurement on a Bose-condensed state effectively ``collapses'' 
the state to a single Bose-Einstein condensate. The Bose-condensed
state before the measurement is a quantum superposition over ideal 
condensates $\int {\cal D}\phi~\psi(\phi)~|N:\phi\rangle$ with different 
condensate wave functions $\phi(\vec x)$ but as position.(\ref{QQU1vv},\ref{QQU1vu}) s of more and more
particles in the state are measured the state of remaining particles
gradually collapses towards one of the ideal condensates $|N:\phi\rangle$.
As a result the density measurement outcome is 
$\rho(\vec x|\phi)=N|\phi(\vec x)|^2$ with probability  
$P(\phi)\approx|\psi(\phi)|^2$.

In our recent paper \cite{paper} we developed a measurement theory
predicting probability of different density measurement outcomes
on a Bose-Einstein condensate evolving under external time-dependent
perturbation. At zero temperature in the framework of the Bogoliubov
theory the state of the condensate is described by a time-dependent
dynamical Bogoliubov vacuum. The dynamical vacuum has simple diagonal
form which leads to a simple gaussian probability distributions
for different density measurement outcomes. The measurement theory
for dynamical Bogoliubov vacuum is important because in many experiments, 
like phase imprinting of dark solitons \cite{Hannover} or condensate splitting 
in atom interferometers \cite{interferometer}, manipulation of the condensate 
generates substantial dynamical depletion which can qualitatively affect 
measured density patterns.

\section{ Diagonal Bogoliubov vacuum }

According to Bogoliubov theory with well-defined number of atoms 
\cite{CastinDum} a condensate which was initially
prepared in its ground state (i.e. a stationary Bogoliubov vacuum)
evolves into a time-dependent excited state which is formally a time-dependent
or dynamical Bogoliubov vacuum annihilated by time-dependent Bogoliubov 
quasiparticle annihilation operators. The dynamical Bogoliubov vacuum can be 
brought to the diagonal form \cite{paper}
\be
|0_b\rangle~\sim~
\left(
\hat a_0^\dagger \hat a_0^\dagger+
\sum_{\alpha=1}^M
\lambda_\alpha
\hat a_\alpha^\dagger \hat a_\alpha^\dagger
\right)^{N/2}~
|0\rangle~,
\label{0lambda}
\ee
with a time-dependent orthonormal basis of single particle states 
$\phi_\alpha(t,\vec x)$ and real time-dependent eigenvalues 
$\lambda_\alpha(t)\in[0,1)$. In the diagonal state (\ref{0lambda})
we sum over finite number $M$ of non-condensate modes, keeping in mind
that $M\to\infty$. However, in practical calculations which almost
always involve some numerics one is forced to work with a finite number
of modes. This is why in the following we will keep finite $M$ in all summations
without any further comment.
As the diagonal vacuum was derived elsewhere 
\cite{paper}, here we only mention that $\phi_0(t,\vec x)$ is a condensate
wave function solving time-dependent Gross-Pitaevskii equation and 
$\phi_\alpha(t,\vec x)$'s are eigenmodes of the density matrix
\bea
&&
\sum_{m=1}^M u^*_m(t,\vec x)u_m(t,\vec y)~=~
\nonumber\\
&&
\sum_{\alpha=1}^M 
[1+{\rm d}N_\alpha(t)]~
\phi^*_\alpha(t,\vec x)\phi_\alpha(t,\vec y)~.
\label{uu}
\eea
Here $[u_m(t,\vec x),v_m(t,\vec x)]$'s are Bogoliubov modes solving 
time-dependent Bogoliubov-de Gennes equations. 
In the number-conserving Bogoliubov theory modes $u_m(t,\vec x)$ and
$v^*_m(t,\vec x)$ are orthogonal to the condensate wave function
$\phi_0(t,\vec x)$. ${\rm d}N_\alpha(t)$ is 
average number of atoms depleted from the condensate wave function 
$\phi_0(t,\vec x)$ to the mode $\phi_\alpha(t,\vec x)$. The eigenvalues 
$\lambda_\alpha(t)$ in the diagonal vacuum (\ref{0lambda}) are 
\be
\lambda_\alpha~=~\sqrt{\frac{{\rm d}N_\alpha}{1+{\rm d}N_\alpha}}~.
\label{lambda}
\ee
Phases of the modes $\phi_\alpha(t,\vec x)$ are chosen in such a way that
the eigenvalues of the matrix 
\bea
&&
\sum_{m=1}^M v_m^*(t,\vec x)u_m(t,\vec y)~=~
\nonumber\\
&&
\sum_{\alpha=1}^M 
\sqrt{{\rm d}N_\alpha(t)[1+{\rm d}N_\alpha(t)]}~
\phi_\alpha(t,\vec x)\phi_\alpha(t,\vec y)~.
\label{vu}
\eea
are real and positive. For the sake of completeness we also quote here
the diagonal density matrix for non-condensate modes
\bea
&&
\sum_{m=1}^M 
v_m(t,\vec x)v_m^*(t,\vec y)~=~
\nonumber\\
&&
\sum_{\alpha=1}^M 
{\rm d}N_\alpha(t)~
\phi_\alpha^*(t,\vec x)\phi_\alpha(t,\vec y)~.
\label{vv}
\eea
One of the results of Ref.\cite{paper} is that the operators 
(\ref{uu},\ref{vu},\ref{vv}) can be simultaneously diagonalized by the
same single particle non-condensate modes $\phi_\alpha(t,\vec x)$. Phases 
of these modes are uniquely defined by the requirement that the eigenvalues
of the operator (\ref{vu}) are real and positive.

\section{Superposition over condensates}

Having in mind applications in quantum measurement theory it is useful
to rewrite the diagonal state (\ref{0lambda}) as a gaussian superposition 
over $N$-particle condensates
\begin{eqnarray}
|0_b\rangle~\sim~
\int dq~
e^{-\sum_{\alpha=1}^M
    \frac{1-\lambda_\alpha}{2\lambda_\alpha}
    q_\alpha^2 }~
|N:\phi(\vec x|q)\rangle~.
\label{0q}
\end{eqnarray}
with the normalized condensate wave functions
\begin{equation}
\phi(\vec x|q)~=~
\frac{ \phi_0(\vec x)+
       \frac{1}{\sqrt{N}}\sum_{\alpha=1}^M 
       q_\alpha\phi_\alpha(\vec x) }
     { \sqrt{1+\frac{1}{N}\sum_{\beta=1}^M q_\beta^2} }~.
\label{phiq}
\end{equation}
Here the state $|N:\phi(\vec x|q)\rangle$ is perfect condensate with
all $N$ particles in the same condensate wave function $\phi(\vec x|q)$.
The (rather technical) proof of the equivalence between the diagonal vacuum in
Eq.(\ref{0lambda}) and the gaussian superposition in Eq.(\ref{phiq}) is
given in Ref.\cite{paper}. The equivalence is approximate, it requires that 
the average number of depleted particles is much less than the total number of 
particles,
\be
{\rm d}N~\equiv~
\sum_{\alpha=1}^M {\rm d}N_\alpha ~\ll~
N~.
\ee
This is not a new assumption but the usual requirement in the Bogoliubov theory.

\section{ Connection with the symmetry breaking approach }

This section is a brief digression on the standard $U(1)$ symmetry breaking 
version 
of the Bogoliubov theory where one splits the field operator into a $c$-number
condensate part plus a small quantum fluctuation
\be
\hat\Psi(\vec x)~=~\sqrt{N}~\phi_0(\vec x)~+~\hat\psi(\vec x)~.
\ee
The quantum fluctuation is further Bogoliubov transformed as
\be
\hat\psi(\vec x)~=~
\sum_{m=1}^M 
\left[~\hat b_m~\tilde{u}_m(\vec x)~+~
       \hat b_m^\dagger~\tilde{v}_m^*(\vec x)~\right]~.
\ee
The initial stationary Bogoliubov vacuum state evolves into a time-dependent
Bogoliubov vacuum state annihilated by the operators $b_m$. In the $U(1)$ 
symmetry breaking Bogoliubov theory the reduced single particle density matrix 
in the time-dependent vacuum state is
\bea
\langle \hat\psi^\dagger(\vec x)\hat\psi(\vec y) \rangle =
\sum_{m=1}^M 
\tilde{v}_m(t,\vec x)\tilde{v}_m^*(t,\vec y)~
\label{U1vv}
\eea
and the anomalous density matrix is
\bea
\langle \hat\psi(\vec x) \hat\psi(\vec y) \rangle =
\sum_{m=1}^M \tilde{u}_m(t,\vec x) \tilde{v}^*_m(t,\vec y)~.
\label{U1vu}
\eea
However, these density matrices are not confined to the subspace orthogonal
to the condensate wave function $\phi_0(t,\vec x)$ because, unlike in the
number-conserving theory, here the Bogoliubov modes $\tilde{u}_m(t,\vec x)$
and $v^*_m(t,\vec x)$ are not exactly orthogonal to $\phi_0(t,\vec x)$. 
This error can be corrected using the relations
\bea
u_m(\vec x)&=&
\tilde{u}_m(\vec x)-\phi_0(\vec x)\langle\phi_0|\tilde{u}_m\rangle
\equiv Q_{\vec x}~ u_m(\vec x)~,
\nonumber\\
v_m(\vec x)&=&
\tilde{v}_m(\vec x)-\phi_0^*(\vec x)\langle\phi_0^*|\tilde{v}_m\rangle
\equiv Q^*_{\vec x}~ v_m(\vec x)~
\label{projections}
\eea
between Bogoliubov modes in the number-conserving and the symmetry-breaking
theories \cite{CastinDum}. After the projection on the subspace orthogonal
to $\phi_0(t,\vec x)$ we obtain the correct non-condensate density matrix
\bea
&&
Q^*_{\vec x}~Q_{\vec y}~
\langle \hat\psi^\dagger(\vec x)\hat\psi(\vec y) \rangle~=~
\nonumber\\
&&
\sum_{m=1}^\infty 
v_m(t,\vec x)v_m^*(t,\vec y)~=~
\nonumber\\
&&
\sum_{\alpha=1}^\infty 
{\rm d}N_\alpha(t)~
\phi_\alpha^*(t,\vec x)\phi_\alpha(t,\vec y)~
\label{QQU1vv}
\eea
and the correct anomalous density matrix 
\bea
&&
Q_{\vec x}~Q_{\vec y}~
\langle \hat\psi(\vec x) \hat\psi(\vec y) \rangle~=~
\nonumber\\
&&
\sum_{m=1}^\infty u_m(t,\vec x) v^*_m(t,\vec y)~=~
\nonumber\\
&&
\sum_{\alpha=1}^\infty 
\sqrt{{\rm d}N_\alpha(t)[1+{\rm d}N_\alpha(t)]}~
\phi_\alpha(t,\vec x)\phi_\alpha(t,\vec y)~.
\label{QQU1vu}
\eea
The last equalities in Eqs.(\ref{QQU1vv},\ref{QQU1vu}) follow from
equations (\ref{vv},\ref{vu}) in the number-conserving theory.

Results of the conventional $U(1)$ symmetry breaking theory can be 
translated into correct results of the number-conserving theory either by

\bi 

\item projecting the Bogoliubov modes on the subspace orthogonal to the
condensate wave function like in Eqs.(\ref{projections}), or

\item projecting directly the density matrices 
$\langle \hat\psi^\dagger(\vec x)\hat\psi(\vec y) \rangle$ and 
$\langle \hat\psi(\vec x) \hat\psi(\vec y) \rangle$ in Eqs. (\ref{U1vv},\ref{U1vu}) 
in a way described by Eqs.(\ref{QQU1vv},\ref{QQU1vu}),

\ei
or any other way to elliminate contamination by the condensate wave function 
$\phi_0(t,\vec x)$ of what is wrongly believed to be purely non-condensate 
density matrices. 

\section{ Bogoliubov representation }

We have shown that in the limit of large $N$ the two forms (\ref{0lambda})
and (\ref{0q}) of the Bogoliubov vacuum are equivalent. However, there is 
yet another representation of the vacuum which may appeal more to 
some of the readers because it is very similar to the coherent state 
representation of the vacuum in quantum optics. The representation is 
\cite{vortex}
\be
|0_b\rangle~\sim~
\int d^2b~ 
e^{-\frac12b^*b}~
\left|
N:\phi(\vec x|b)
\right\rangle~.
\label{0b}
\ee
Here $d^2b=\prod_{m=1}^Md^2b_m$, $b^*b=\sum_{m=1}^Mb_m^*b_m$ and
$\phi(\vec x|b)$ is a normalized condensate wave function
\bea
&&
\phi(\vec x|b)=  \\
&&
{\cal N}
\left\{
       \phi_0(t,\vec x)+
       \frac{1}{\sqrt{N}}
       \sum_{m=1}^M
       \left[  
       b_m u_m(t,\vec x)+
       b_m^* v_m^*(t,\vec x)
       \right]                  
\right\}
\nonumber
\label{phib}
\eea
where the normalization factor is
\be
{\cal N}^{-1}=
1+
\frac{1}{N}
\sum_{m,n=1}^M
\left\langle
b_m u_m + b_m^* v_m^*|
b_n u_n + b_n^* v_n^*
\right\rangle~.
\label{calN}
\ee
In this Section we will prove
equivalence between Eq.(\ref{0b}) and Eq.(\ref{0q}) in the limit of large $N$. 
The main difference between these equations is that in Eq.(\ref{0q}) there are $M$ 
integrations over {\it real} coordinates $q_\alpha$ while in Eq.(\ref{0b}) there 
are also $M$ integrations but over {\it complex} coordinates $b_m$. In this sense 
Eq.(\ref{0q})is more compact representation than Eq.(\ref{0b}).

As a first step we use the large $N$ limit to transfer the normalization factor 
${\cal N}$ to the exponent in Eq.(\ref{0b}),
\bea
\int &d^2b&~ 
e^{-\frac12\sum_m b^*_mb_m
   -\frac12\sum_{mn}\langle b_mu_m+b_m^*v_m^* |
                            b_nu_n+b_n^*v_n^*  \rangle}~
\nonumber\\
&&
~
\left(
\hat a_0^\dagger+
\frac{1}{\sqrt{N}}
\sum_{m=1}^M
\left[
b_m   \hat u_m^\dagger+
b_m^* \hat v_m^\dagger
\right]
\right)^N
|0\rangle~.
\label{0b+}
\eea
Here the new operators are $\hat u_m=\langle u_m|\hat\psi\rangle$ and
$\hat v_m=\langle v_m^*|\hat\psi\rangle$. At this point we make a
transformation to the orthonormal basis of $\phi_\alpha$'s,
$
\sum_{m=1}^M
\left[
b_m \hat u_m^\dagger + b_m^* \hat v_m^\dagger =
\right]
\sum_{\alpha=1}^M
z_\alpha
\hat a_\alpha^\dagger
$,
or equivalently 
\be
\sum_m \left[ b_mu_m+b_m^*v_m^* \right]=\sum_\alpha z_\alpha\phi_\alpha~.
\label{bz}
\ee
After this transformation the state (\ref{0b+}) becomes
\bea
\int &d^2z&~ 
e^{-\frac12\sum_m b^*_mb_m
   -\frac12\sum_\alpha z_\alpha^*z_\alpha}~
\nonumber\\
&&
~
\left(
\hat a_0^\dagger+
\frac{1}{\sqrt{N}}
\sum_{\alpha=1}^M
z_\alpha \hat a_\alpha^\dagger
\right)^N
|0\rangle~.
\label{0b++}
\eea
Here $b$'s are linear functions of $z$'s
\be
b_m=\sum_{\alpha=1}^M 
\left[
z_\alpha   \langle u_m|\phi_\alpha\rangle-
z_\alpha^* \langle v_m|\phi_\alpha^*\rangle
\right]
\label{zb}
\ee 
obtained by inverting the transformation (\ref{bz}). Using this inverse 
transformation and equations (\ref{uu},\ref{vu},\ref{vv}) we rewrite the 
exponent in Eq.(\ref{0b++}) as
\bea
&& \sum_m b^*_mb_m+\sum_\alpha z_\alpha^*z_\alpha~=
\nonumber\\
&& \sum_\alpha 
2(1+{\rm d}N_\alpha)z_\alpha^*z_\alpha -
\nonumber\\
&&
\sqrt{{\rm d}N_\alpha(1+{\rm d}N_\alpha)} z_\alpha^*z_\alpha^* -
\sqrt{{\rm d}N_\alpha(1+{\rm d}N_\alpha)} z_\alpha  z_\alpha~.
\label{exp}
\eea
With this exponent the gaussian integral in Eq.(\ref{0b++}) gives
a simple correlator for $z$'s
\bea
\int & d^2z & 
e^{-\frac12\sum_m b^*_mb_m
   -\frac12\sum_\alpha z_\alpha^*z_\alpha}~
z_\beta z_\gamma~=~
\nonumber\\
&&
\delta_{\beta\gamma}~\sqrt{\frac{{\rm d}N_\beta}{1+{\rm d}N_\beta}}~=~
\delta_{\beta\gamma}~\lambda_\beta~.
\label{corrz}
\eea
In the last equality we use the relation in Eq.(\ref{lambda}).
The gaussian state in Eq.(\ref{0b++}) is completely determined
by this correlator. The same correlator for real $q$'s is also 
obtained in the real gaussian integral
\be
\int dq~
e^{-\frac12\sum_\alpha\frac{q_\alpha^2}{\lambda_\alpha}}~
q_\beta~q_\gamma~=~
\delta_{\beta\gamma}~\lambda_\beta~.
\label{corrq}
\ee
Comparing the correlators (\ref{corrz}) and (\ref{corrq}) we find that
the gaussian state (\ref{0b++}) is equal to the state
\bea
\int dq~ 
e^{-\frac12\sum_\alpha \frac{q_\alpha^2}{\lambda_\alpha} }~
\left(
\hat a_0^\dagger+
\frac{1}{\sqrt{N}}
\sum_{\alpha=1}^M
q_\alpha \hat a_\alpha^\dagger
\right)^N
|0\rangle~
\label{0b+++}
\eea
with real coordinates $q$. Finally after normalization of the creation
operator we get
\bea
\int dq~ 
e^{-\sum_\alpha \frac{(1-\lambda_\alpha)q_\alpha^2}{2\lambda_\alpha} }~
\left(
\frac{ \hat a_0^\dagger+
       \frac{1}{\sqrt{N}}
       \sum_{\alpha=1}^M
       q_\alpha \hat a_\alpha^\dagger  }
     { \sqrt{1+\frac{1}{N}\sum_{\beta=1}^M q_\beta^2} }
\right)^N
|0\rangle~
\label{0b++++}
\eea
and this state is the gaussian representation
of Bogoliubov vacuum in Eq.(\ref{0q}). In this way we have shown that
the representations (\ref{0q}) and (\ref{0b}) are equivalent. 

The complex representation (\ref{0b}) with Bogoliubov coefficients $b_m$ 
has twice as many integration variables as the real representation (\ref{0q}) 
with real coordinates $q_\alpha$. What is more, the real representation 
(\ref{0q}) is spanned by the orthonormal basis of $\phi_\alpha$'s while the 
complex representation is spanned by the Bogoliubov modes $(u_m,v_m)$ which 
are not orthogonal in the usual sense. The real representation reveals 
the diagonal structure of the vacuum which is implicit in the complex
representation. 

\section{ Probability distribution }\label{prob}

Having shown equivalence between different representations of the dynamical 
Bogoliubov
vacuum (\ref{0lambda},\ref{0q},\ref{0b}) we now return to the main subject of
this paper which is probability distribution for different condensate 
wave functions. The quantum superposition in the real representation (\ref{0q})
directly leads to the probability distribution for different $q$'s 
\be
P(q)\propto
\prod_{\alpha=1}^M e^{-\frac{q_\alpha^2}{2{\rm d}N_\alpha}}~
\label{Pq}
\ee 
which is valid for the non-condensate modes with large occupation numbers
${\rm d}N_\alpha$ when we can replace 
\be
\frac{1-\lambda_\alpha}{2\lambda_\alpha}\approx
\frac{1}{4{\rm d}N_\alpha}~,
\ee
compare Eq.(\ref{lambda}). The distribution $P(q)$ is the desired
probability distribution for different possible density measurement 
outcomes given by
\bea
&&
\rho(\vec x|q)~=~N|\phi(\vec x|q)|^2~=\nonumber\\
&&
N\left|
\frac{\phi_0(t,\vec x)+\frac{1}{\sqrt{N}}\sum_{\alpha=1}^Mq_\alpha\phi_\alpha(t,\vec x)}
     {\sqrt{1+\frac{1}{N}\sum_{\beta=1}^N q_\beta^2}}
\right|^2~.
\label{rhoq}
\eea
Equations (\ref{Pq}) and (\ref{rhoq}) define a simple scheme how to simulate
possible density measurement outcomes on the dynamical Bogoliubov vacuum state. 

\section{Measurements in momentum space}

Up to this point we assumed ideal density measurement where all particle
positions are measured at the same time. However, there is no reason
why these results should be limited to measurements of particle momenta
instead of position. In fact the diagonal second quantized
representation (\ref{0lambda}) prefers neither position nor momentum 
representation, and consequently all the following formulas can be rewritten
in momentum representation simply by replacing $\vec x \to \vec k$ and
$\vec y \to \vec p$. In the end we arrive at the probability distribution
$P(q)$ in Eq.(\ref{Pq}) for different possible outcomes of momentum 
distribution
\bea
\rho(\vec k|q)=
N\left|
\frac{\phi_0(t,\vec k)+\frac{1}{\sqrt{N}}\sum_{\alpha=1}^Mq_\alpha\phi_\alpha(t,\vec k)}
     {\sqrt{1+\frac{1}{N}\sum_{\beta=1}^N q_\beta^2}}
\right|^2~.
\label{rhoqk}
\eea
A quantum measurement on the dynamical Bogoliubov vacuum state which measures 
all particle momenta at the same time will give momentum density distributions
(\ref{rhoqk}) with probabilities (\ref{Pq}).

\section{Conclusion}

In conclusion, we derived a convenient diagonal form of the time-dependent
Bogoliubov vacuum which greatly facilitates simulations of quantum measurements
on Bose-condensed systems both in position and momentum space.

{\bf Acknowledgements. --- } Work of JD was supported in part by the KBN 
grant PBZ-MIN-008/P03/2003 and work of KS was supported in part by Polish 
Government scientific funds (2005-2008) as a research project. 

\end{document}